\documentclass[conference]{IEEEtran}
\usepackage{comment}
\usepackage{multicol}
\usepackage{blindtext}
\usepackage{amssymb}
\usepackage{subfigure}
\usepackage{amsmath}
\usepackage{amsfonts}
\usepackage{mathrsfs}
\usepackage{engord}
\usepackage[dvips]{graphicx}
\usepackage{bbm}
\usepackage{threeparttable}
\usepackage{booktabs}
\usepackage{epsfig}
\usepackage{color}
\usepackage{graphicx}
\usepackage{multirow}
\usepackage{cite}
\usepackage{epstopdf}
\usepackage{amsthm}

\usepackage{framed}
\usepackage{url}
\usepackage[ruled,norelsize]{algorithm2e}

\makeatletter
\newcommand{\removelatexerror}{\let\@latex@error\@gobble}
\makeatother
\begin{document}

\newtheorem{thm}{Theorem}
\newtheorem{prop}{Proposition}
\newtheorem{reproof}{Proof}
\newtheorem{lem}{Lemma}
\newtheorem{defn}{Definition}
\newtheorem{ex}{Example}
\newtheorem{cor}{Corollary}
\newtheorem{prn}{Principle}
\newtheorem{case}{Case}
\title{Latent Factor Analysis of Gaussian Distributions under  Graphical Constraints}

\author{\IEEEauthorblockN{Md Mahmudul Hasan,
Shuangqing Wei, Ali Moharrer}}
\maketitle
\footnotetext[1]{Md M Hasan,   S. Wei and A. Moharrer  are with the school of Electrical Engineering and Computer Science, Louisiana State University, Baton Rouge, LA 70803, USA (Email: mhasa15@lsu.edu, swei@lsu.edu, alimoharrer@gmail.com). }



\begin{abstract}
The Constrained Minimum Determinant Factor
Analysis (CMDFA) setting was motivated by Wyner's common information problem where we seek a latent representation of a given Gaussian vector distribution with the minimum mutual information under certain generative constraints. In this paper, we explore the algebraic structures of the solution space of the CMDFA, when the underlying covariance matrix $\Sigma_x$ has an additional latent graphical constraint, namely, a latent star topology. 
In particular, sufficient and necessary conditions in terms of the relationships between edge weights of the star graph  have been found. Under such conditions and constraints, we have shown that the CMDFA problem has either a rank one solution or a rank $n-1$ solution where $n$ is the dimension of the observable vector.  Further results are given in regards to the solution to the CMDFA with $n-1$ latent factors. 
\end{abstract}
\section{INTRODUCTION}
Factor Analysis (FA) is a commonly used tool in multivariate statistics to represent the correlation structure of a set of observables in terms of significantly smaller number of variables called ``latent factors". With the growing use in data mining, high dimensional data  analytics, factor analysis has already become a prolific area of research \cite{chen2017structured}\cite{bertsimas2017certifiably}. Classical Factor Analysis models seek to decompose the correlation matrix of an $n$-dimensional  random vector ${\bf X} \in {\mathcal R}^n$, $\Sigma_x  $, as the sum of a diagonal matrix $ D $ and a Gramian matrix $ \Sigma_{x}-D $. 

The literature that approached Factor Analysis can be classified in three major categories. Firstly, algebraic approaches \cite{albert1944matrices} and  \cite{drton2007algebraic}, where the principal aim was to give a characterization of the vanishing ideal of the set of symmetric $ n\times n $ matrices that decompose as the sum of a diagonal matrix and a low rank matrix, did not offer scalable algorithms for higher dimensional statistics. Secondly, Factor Analysis  via heuristic local optimization techniques, often based on the expectation maximization algorithm, were computationally tractable  but offered no provable performance guarantees. The third and final type are the  convex optimization based methods such as Constrained Minimum Trace Factor Analysis (CMTFA) \cite{bentler1980inequalities} \cite{} and CMDFA \cite{moharrer2017algebraic}.  The motivation behind CMDFA comes from Wyner's common information $ C(X_{1},X_{2}) $ which characterizes  the minimum amount of common randomness needed to approximate the joint density between a pair of random variables $ X_{1} $  and $ X_{2} $ to be $C(X_{1},X_{2})= \min_{\underset{X_{1}-Y-X_{2}}{P_{Y}} } I(X_{1},X_{2}; Y) $, where $ I(X_{1},X_{2}; Y) $ is the mutual information between $ X_{1} $, $ X_{2} $ and $ Y $, $ X_{1}-Y-X_{2} $ indicates the conditional independence between $ X_{1} $ and $ X_{2} $ given $ Y $, and the joint density function is sought to esnure such conditional independence as well as the given joint density of $ X_{1} $  and $ X_{2} $. Since the Factor Analysis of the  Gaussian random vector $ \vec{X} $  can be modelled as $\vec{X}=A\vec{Y}+ \vec{Z}$,
where $A_{n\times k}$ is a real matrix, $ \vec{Y}_{k\times 1}, k<n $ is the vector of independent latent variables   and $\vec{Z}_{n\times 1}$ is a Gaussian vector of zero mean and covariance matrix $\Sigma_z = D$. Hence we have, 
$I(\vec{X};\vec{Y})=h(\vec{X})-h(\vec{X}|\vec{Y})=h(\vec{X})-h(\vec{Z})$
where $ I(\vec{X};\vec{Y}) $ is the mutual information between $ \vec{X} $ and $ \vec{Y} $, $ h(\vec{X}), h(\vec{Z})  $ are differential entropies of $ \vec{X}$ and $ \vec{Z} $ and $ h(\vec{X}|\vec{Y}) $ is the conditional entrophy of $ \vec{X} $ given $ \vec{Y} $.  Hence characterizing the common information between $ \vec{X} $ and $ \vec{Y} $ \cite{xu2016lossy}\cite{wyner1975common}\cite{satpathy2015gaussian}  would be $ \min_{A,\Sigma_{z}} I(\vec{X};\vec{Y}) $ which is an equivalent problem to $ \max_{\Sigma_{z}} h(\vec{Z})$ hence equivalent to $ \min_{\Sigma_{z}} -\log |\Sigma_{z}| $. 

The scope of this paper is limited to analysing the solution space of CMDFA and recovering the underlying graphical structures.  It is important to remark that our work is not concerned about the algorithm side of the optimization technique, rather our focus is to characterize and find insights about their solution space. Moharrer and Wei \cite{moharrer2017algebraic} derived CMDFA from a broader class of convex optimization problem and established  a relationship between the outcome of these optimization techniques and  common information problem \cite{wyner1975common}.  We find the explicit conditions under which the CMDFA solution of $ \Sigma_{x} $ recoves a star structure. Since star may not always be the optimum solution, we have also shown the existence and uniqueness of a rank $ n-1 $ CMDFA solution of $ \Sigma_{x} $ which is the only other possible solution. We have shown analyticaly the optimality of the non-star solution over star under certain circumstances from common information point of view, and at the end presened some numerical data to support our claims. 
\section{Definitions and Notations}
Let $ \vec{b} $ be a real $ n $ dimensional column vector  and $ A $ be an $ n\times n $ matrix. As in literature in general we denote the $ i $th element $ \vec{b} $ as $ b_{i} $ and the $ (i,j) $th element of $ A $ as $ A_{i,j} $. Here we define all the vector operations and notations in terms of  $ \vec{b} $ and $ A $, that will carry their meaning on other vectors and  matrices throughout this paper unless stated otherwise. 

 Let $ \vec{a}_{i,*} $ and $  \vec{a}_{*,i}  $ denote the $i$th row and $ i $th column vector of matrix $ A $ respectively. Function $ \lambda_{min}(A) $ is defined to be the smallest eigen value of matrix $ A $. $ N(A) $ stands for the null space of matrix $ A $. 

Vectors $ \vec{1} $ and $ \vec{0} $ are the $ n $ dimensional column vectors with each element equal to $ 1 $ and $ 0 $ respectively. When we write $ \vec{b}\geq 0 $ we mean that each element of the vector $ b(i)\geq 0, 1\leq i \leq n $.  $ \vec{b}^{2} $ is the Hadamard product of vector $ \vec{b} $ with itself. $ ||\vec{b} ||$ denotes the $ L_{2} $ norm of vector $ \vec{b} $. 

Now we define two terms i.e. \textit{dominance} and \textit{non-dominance} of a vector which will repeatedly appear throughout the paper. When we talk about the dominance or non-dominance of any vector $ \vec{b} $ we assume that the elements of the vector are sorted in a way such that $ |b_{1}|\geq |b_{2}|\geq \dots \geq |b_{n}|$. We call vector $ \vec{b} $ dominant and $ b_{1} $ the dominant element if for the above sorted vector $ |b_{1}|>\sum_{j\neq 1} |b_{j}| $  holds.  Otherwise $ \vec{b} $ is non-dominant.
\section{Formulation of the Problem} 
First of all we define the real column vector $ \vec{\alpha} $ as $\vec{\alpha} = [\alpha_1, \dots, \alpha_n]' \in {\mathcal R}^n$ where  $ 0< |\alpha_{j}|< 1 $, $ j= 1,2, \dots ,n $ and 
\begin{align}\label{order_alpha}
|\alpha_{1}|\geq |\alpha_{2}| \geq \dots \geq |\alpha_{n}|  
\end{align}
Let us consider a star structured population covariance matrix $ \Sigma_{x} $ having all the diagonal comptonents $ 1 $  as given by equation \eqref{sigmax}.
 \begin{equation}\label{sigmax}
\Sigma_{x}=\begin{bmatrix}
1&\alpha_{1}\alpha_{2}&\dots&\alpha_{1}\alpha_{n}\\
\alpha_{2}\alpha_{1}&1&\dots&\alpha_{2}\alpha_{n}\\
\vdots&\vdots&\ddots&\vdots\\
\alpha_{n}\alpha_{1}&\alpha_{n}\alpha_{2}&\dots&1\\
\end{bmatrix}
\end{equation}
Traditional Factor Analysis problems seeks to decompose $ \Sigma_{x} $ as the sum of a low rank (rank $ <n $ ) component and a diagonal matrix. This problem of finding a low rank solution for the decomposition of $ \Sigma_{x} $ under certain constrains has been equivalently formulated as a particular class of convex optimization problem in \cite{della1982minimum}. CMDFA aims to minimize the mutual information between the observable Gaussian random vector $X$ and the latent ones $Y$. It is thus to seek joint distribution between the latent and observable ones such that the conditional entropy $H(X|Y)$ is maximized. Under the joint Gaussian distribution, it is the same as seeking factorization of $\Sigma_X$ such that the determinant of the matrix $D$ is minimized as in equation \eqref{decomposition} under the constraint that both $ (\Sigma_{x}-D) $ and $ D $ are Gramian matrices.
\begin{align}\label{decomposition}
\Sigma_{x}=(\Sigma_{x}-D)+D
\end{align}
Though the motivation behind our work is the graphical tree structure given by Figure \ref{figtree}, we limit the scope of the paper to the analysis of one particular node of the tree given by Figure \ref{figstar}. As we can see, a set of observables $ X_{1}, \dots , X_{n} $ connected to a common latent variable $ Y $ makes it a star topology indicating the  conditional independence given by \eqref{conditional independence} among the observables.
\begin{align}\label{conditional independence}
p(X_{1}X_{2},\dots, X_{n}|Y)=\Pi_{i=1}^{n}p(X_{i}|Y)
\end{align} 
We define $ \Sigma_{t} $ as $ \Sigma_{t}=\Sigma_{x} -D$. One latent variable producing $ n $ observables corresponds to matrix $ \Sigma_{t} $ in  equation \eqref{decomposition} having the rank $ 1 $  solution given by,
\begin{align}\label{sigmatnd}
\Sigma_{t,ND}=\begin{bmatrix}
\alpha_{1}^{2}&\alpha_{1}\alpha_{2}&\dots&\alpha_{1}\alpha_{n}\\
\alpha_{2}\alpha_{1}&\alpha_{2}^{2}&\dots&\alpha_{2}\alpha_{n}\\
\vdots&\vdots&\ddots&\vdots\\
\alpha_{n}\alpha_{1}&\alpha_{n}\alpha_{2}&\dots&\alpha_{n}^{2}\\
\end{bmatrix}
\end{align}
or equivalently $ \Sigma_{x} $ being produced by the following graphical model. 
\begin{align}
& \begin{bmatrix}
X_{1}\\
\vdots\\
X_{n}
\end{bmatrix}
=\begin{bmatrix}
\alpha_{1}\\
\vdots\\
\alpha_{n}
\end{bmatrix}
Y
+
\begin{bmatrix}
Z_{1}\\
\vdots\\
Z_{n}
\end{bmatrix}\label{graphical model}\\
\Rightarrow &\vec{X}= \vec{\alpha}Y+\vec{Z}
\end{align}
where
\begin{itemize}
\item $ \{X_{1}, ..., X_{n}\}$  are conditionally independent Gaussian random variables given $ Y $, forming the jointly Gaussian random vector $ \vec{X}\sim \mathcal{N}(0,\Sigma_{x}) $ where $ Y\sim \mathcal{N}(0,1) $.
\item $ \{Z_{1}, ..., Z_{n} \}$ are independent Gausian random varables with $Z_{j}\sim \mathcal{N}(0,1-\alpha_{j}^{2})\quad 1\leq j \leq n $ forming the Gaussian random vector $ \vec{Z} $.
\end{itemize}
The aforementioned convex optimization problem CMDFA seeks low rank solution to equation \eqref{decomposition} but not necessarily a rank $ 1 $ solution. It remains to be seen if  CMDFA solution to $ \Sigma_{x} $ recovers the graphical model given by \eqref{graphical model}. Also to be investigated is the exact solution to CMDFA if it fails to recover the underlying star topology. In the rest of the paper, we will present both sufficient and necessary conditions in terms of the relationships between $\theta_j$  under which the rank of the optimal $\Sigma_t$ and the values of $D$'s entries are determined. 

\section{Solutions to CMDFA}
In this section we present the detailed analysis  of the CMDFA solution space of $ \Sigma_{x} $. We defne the real column vector $ \vec{\theta} \in {\mathcal R}^n$ as $\vec{\theta}=[\theta_{1},\dots, \theta_{n} ]'$ where $ \theta_{i}=\frac{|\alpha_{i}|}{\sqrt{1-\alpha_{i}^{2}}}, 1\leq i\leq n $.
 
As we can see, each elements in $ \vec{\theta} $ is equal to  the square root of the signal to noise ratio ($ \sqrt{\text{SNR}} $) of the corresponding element of vector $ \vec{\alpha} $. The following order of the elements of  $ \vec{\theta} $  is a necessary consequence of our assumption in \eqref{order_alpha},
\begin{align}\label{order_theta}
|\theta_{1}|\geq |\theta_{2}| \geq \dots \geq |\theta_{n}|
\end{align}
As we metioned  before, we are interested to find out if CMDFA low rank decomposition of $ \Sigma_{x} $ produces a rank $ 1 $ matrix. Next we analyse the solution space of CMDFA and find explicit conditions for both when the solution is rank $ 1 $ and when it is not. To start the proceedings we state Theorem \ref{sufficient and necessary cmdfa} given in \cite{moharrer2017algebraic} that gives
 the necessary and sufficient condition for $ D^{*}$ to be the CMDFA solution of the decomposition given in \eqref{decomposition}. 
\begin{thm}\label{sufficient and necessary cmdfa}
The matrix  $ D^{*}$ is the CMDFA solution of $ \Sigma_{x} $ if and only if $ \lambda_{min}( \Sigma_{x} -D^{*})=0 $, and there exists $ n\times k $ matrix $ T $ such that  $\vec{t}_{*,i}\in \mathcal{N}(\Sigma_{x}-D^{*}), 1\leq i \leq k$ and  $||\vec{t}_{i,*}||^{2}=\frac{1}{1-\alpha_{i}^{2}}, 1\leq i \leq n$.
\end{thm}

In the first of the two subsections of this section, we find  the conditions under which CMDFA solution of $ \Sigma_{x} $  recovers the model given by \eqref{graphical model} or equivalently speaking, find condtions under which CMDFA solution of $ \Sigma_{x} $ is the rank $ 1 $ matrix given by \eqref{sigmatnd}.   In the other subsection, we show the detailed analysis on the existance and uniqueness of the CMDFA solution of $ \Sigma_{x} $, when the solution is not a rank $ 1 $ matrix.
\subsection{CMDFA Non-dominant Case}
Here we analyse the conditions under which the CMDFA solution of $ \Sigma_{x} $ recovers a star structure. Lemma \ref{lem1} sets the groundwork for the Theorem to follow. The Lemma also has a geometric interpretaion that enriches our overall understanding of the CMDFA non-dominace case. 
\begin{lem}\label{lem1}
There exists matrix $ T $ of dimension $ n\times k $ such that $ \vec{t}_{*,i}\in N(\Sigma_{t,ND}), \quad 1\leq i \leq k $  and $ ||\vec{t}_{j,*}||^{2}=\frac{1}{1-\alpha_{j}^{2}}, \quad 1\leq j \leq n $ if and only if  vector $ \vec{\theta} $ is non-dominant i.e.,
\begin{align}\label{non dominance of theta}
\theta_{1}\leq \sum_{i=2}^{n}\theta_{i}
\end{align}
\end{lem}

The proof of Lemma \ref{lem1} and the associated geometric interpretation is given in \cite{}. We are now well equipped to state and prove the statement of Theorem \ref{th3} that has the main result of this subsection.
\begin{thm}\label{th3}
CMDFA solution of $ \Sigma_{x} $ is $ \Sigma_{t,ND} $ if and only if $ \vec{\theta} $ is non-dominant. 
\end{thm}
The theorem states that the CMDFA solution to a star connected network is a star itself,  if and only if there is no dominant element in the vector $ \vec{\theta} $.
\begin{proof}[\textbf{Proof of Theorem \ref{th3}:}]
Now  we refer back to the necessary and sufficient condition for CMDFA solution at the begining of this section given by Theorem \ref{sufficient and necessary cmdfa}. Since, $ \Sigma_{t,ND} $ in rank $ 1 $, its minimum eigenvalue is $ 0 $. To complete the proof of Theorem \ref{th3}, we only need to show the existance of matrix $ T$ such that the column vectors of $ T$ are in the null space of $ \Sigma_{t,ND} $ and the  $ L_{2} $-norm square of the $ i$th row of $ T $ is $ \frac{1}{1-\alpha_{i}^{2}}, 1\leq i \leq n $. 

Lemma \ref{lem1} has already shown that, for the existence of such $ T $ non-dominance given by equation \eqref{non dominance of theta} is a necessary condition. Next we show, by constructing such a $ T $ matrix under the assumption of non-dominance of $ \vec{\theta} $, that non-dominace is also a sufficient condition . And that should complete the proof of Theorem \ref{th3}. 

Its trivial to find  the following  basis vectors for the null space of  $ \Sigma_{t,ND} $,

\begin{align}
v_{1}=\begin{bmatrix}
-\frac{\alpha_{2}}{\alpha_{1}}\\
1\\
0\\
\vdots\\
0
\end{bmatrix}, v_{2}=\begin{bmatrix}
-\frac{\alpha_{3}}{\alpha_{1}}\\
0\\
1\\
\vdots\\
0
\end{bmatrix},\dots, \quad v_{n-1}=\begin{bmatrix}
-\frac{\alpha_{n}}{\alpha_{1}}\\
0\\
0\\
\vdots\\
1
\end{bmatrix}
\end{align}

We define matrix $ V $ so that its columns span the null space of $ \Sigma_{t,ND} $,
\begin{align}\label{v}
&V\notag\\
&= \begin{bmatrix}
-\frac{\alpha_{2}}{\alpha_{1}}&\dots&-\frac{\alpha_{n}}{\alpha_{1}}&-\left(c_{2}\frac{\alpha_{2}}{\alpha_{1}}+\dots+c_{n}\frac{\alpha_{n}}{\alpha_{1}}\right)\\
1&\dots&0&c_{2}\\
0&\dots&0&c_{3}\\
\vdots&\ddots&\vdots&\vdots\\
0&\dots&1&c_{n}
\end{bmatrix}
\end{align} 
where $ c_{i}=\frac{\widetilde{c}_{i}}{\sqrt{1-\alpha_{i}^{2}}}, \quad i=2,\dots, n $  and $ \widetilde{c}_{i}\in \{1,-1\} $.

The  columns of $ V $ span the null space of $ \Sigma_{t,ND} $. To construct our desired matrix $ T $, under the assumption of non-dominance of $ \vec{\theta} $, it will suffice for us to find a diagonal matrix $ B_{n \times n} $ such that the following holds. 
\begin{align}\label{t}
T_{n \times n}= V_{n \times n}\cdot B_{n \times n}
\end{align}
where the $ L_{2} $-norm square of the $ i $th row of $ T $ is $ \frac{1}{1-\alpha_{i}^{2}}  $. Using \eqref{t}, 
\begin{align}\label{tt}
TT'= VBB'V'=V \beta V'
\end{align}
We require the diagonal  matrix $ \beta $ to have only non-negative entries. Based on the conditions imposed on the matrix $ T $, we have the following $ n $ equations,

\begin{align}\label{d1}
&\frac{\alpha_{2}^{2}}{\alpha_{1}^{2}}\beta_{11}+\frac{\alpha_{3}^{2}}{\alpha_{1}^{2}}\beta_{22}+\dots+\frac{\alpha_{n}^{2}}{\alpha_{1}^{2}}\beta_{n-1,n-1}+\notag\\
&\left(c_{2}\frac{\alpha_{2}}{\alpha_{1}}+c_{3}\frac{\alpha_{3}}{\alpha_{1}}+\dots+c_{n}\frac{\alpha_{n}}{\alpha_{1}}\right)^{2}\beta_{nn}=\frac{1}{1-\alpha_{1}^{2}}
\end{align}

\begin{align}\label{d2}
\beta_{ii}+c_{i+1}^{2}\beta_{nn}=\frac{1}{1-\alpha_{i+1}^{2}}, \quad i=1,\dots,n-1
\end{align}

Solving, \eqref{d1} with the help of \eqref{d2} we get, 
\begin{align}
\beta_{nn}=
&\frac{\frac{\alpha_{1}^{2}}{1-\alpha_{1}^{2}}-\frac{\alpha_{2}^{2}}{1-\alpha_{2}^{2}}-\dots-\frac{\alpha_{n}^{2}}{1-\alpha_{n}^{2}}}{\sum_{i\neq j, i\neq 1, j\neq 1}c_{i}c_{j}\alpha_{i}\alpha_{j}}
\end{align}
It is straightforward to see that, to ensure all the $ \beta_{•ii} , 1\leq i \leq n$ are non-negative, we need $ \beta_{nn}\leq 1 $. We select $ \widetilde{c}_{i}, 2\leq i \leq n $ such that, 
\begin{align}
c_{i}\alpha_{i}=\frac{\widetilde{c}_{i}\alpha_{i}}{\sqrt{1-\alpha_{i}^{2}}}=\theta_{i}, \quad i=2,\dots, n
\end{align}
Under such selection of $ \widetilde{c}_{i}, 2\leq i \leq n $, $ \beta_{nn} $ becomes,
\begin{align}
\beta_{nn}=\frac{\theta_{1}^{2}-\theta_{2}^{2}-\dots-\theta_{n}^{2}}{\sum_{i\neq j, i\neq 1, j\neq 1}\theta_{i}\theta_{j}}
\end{align}
Now, using the non-dominance assumption given in \eqref{non dominance of theta}, we have
\begin{align}
&\theta_{1}^{2}\leq \left(\sum_{i=2}^{n}\theta_{i}\right)^{2}\notag\\
\Rightarrow &\frac{\theta_{1}^{2}-\sum_{i=2}^{n}\theta_{i}^2}{\sum_{i\neq j, i\neq 1, j\neq 1}\theta_{i}\theta_{j}}\leq 1\label{bc}\\
\Rightarrow & \beta_{nn}\leq 1
\end{align}
Which means non-dominance of vector $ \vec{\theta} $ is a sufficient condition to construct the kind of $ T $ matrix we are looking for. That completes the proof of Theorem \ref{th3}.
\end{proof}
Lemma \ref{lem2} states a very special case which is proved and discussed in detail in \cite{}. 
\begin{lem}\label{lem2}
When the non-dominance condition given in \eqref{non dominance of theta} holds for equality, the null space of $ \Sigma_{t,ND} $ becomes one dimensional.
\end{lem}
The Lemma essentially states that, under the boundary condition i.e. when the non-dominance condition holds for equality the dimension of the null space of $ \Sigma_{t,ND} $ cannot be anything other than $ 1 $ .

%%%===============================================
\subsection{Dominant Case}
%%%===============================================
Having proved that the non-dominance of vector $ \vec{\theta} $ is a sufficient and necessary condition for CMDFA solution of $ \Sigma_{x} $ to recover a star structure, we are left with only the dominant case now i.e.
\begin{align}\label{domcon}
\theta_{1}>\sum_{i=2}^{n}\theta_{i}
\end{align}
Under the above dominant condition we are going  to show the existence of a rank $ n-1 $ solution of $ \Sigma_{x} $. Any solution we find will be unique, because CMDFA is a special type of the broader class of convex optimization problem defined in \cite{}. We still have to satisfy the same sufficient and necessary condtion for the CMDFA solution, that was set by Theorem \ref{th3}. Like the non-dominant case, for the matrix $ D^*$ to be the CMDFA solution of $ \Sigma_{x} $ under the dominant case, the minimum eigen value of the solution matrix has to be $ \lambda(D^*)=0 $ and the $ L_{2} $-norm square of the $ i $th row of the  null space matrix $ T $ has to be $ \frac{1}{1-\alpha_{1}^{2}} $. Since our conjecture for the dominant case is an $n-1  $ rank solution, the null space matrix $ T $ will be rank $ 1 $ i.e. a column vector. Mathematically speaking, we are trying to show the existance of $ 0<a_{i}<1, 1\leq i \leq n $ such that the following orthogonality condition holds.

\begin{align}\label{bigger picture}
\begin{bmatrix}
a_{1}&\alpha_{1}\alpha_{2}&\alpha_{1}\alpha_{3}&\dots&\alpha_{1}\alpha_{n}\\
\alpha_{2}\alpha_{1}&a_{2}&\alpha_{2}\alpha_{3}&\dots&\alpha_{2}\alpha_{n}\\
\vdots&\vdots&\vdots&\ddots&\vdots\\
\alpha_{n}\alpha_{1}&\alpha_{n}\alpha_{2}&\alpha_{n}\alpha_{3}&\dots&a_{n}\\
\end{bmatrix}
\begin{bmatrix}
\frac{c_{1}}{\sqrt{1-a_{1}}}\\
\vdots\\
\vdots\\
\frac{c_{n}}{\sqrt{1-a_{n}}}
\end{bmatrix}=
\begin{bmatrix}
0\\
\vdots\\
\vdots\\
0
\end{bmatrix}
\end{align}
where $ c_{i}\in \{ -1,1\} $. Once we have such $ a_{i}, 1\leq i \leq n$ the $ i $th diagonal element of the CMDFA solution matrix  $ D^*$ under the dominant case will be $ 1-a_{i},  1\leq i \leq n$. The  above orthogonality relationship gives us the following $ n  $ equations. 
\begin{align}\label{es}
&\frac{a_{i}c_{i}}{\sqrt{1-a_{i}}}+\sum_{j\neq i}\frac{\alpha_{i}\alpha_{j}c_{j}}{\sqrt{1-a_{j}}}=0, 1\leq i \leq n
\end{align}
Let $ (i) $ denote the $ i $th equation given by \eqref{es}. Using the linear combination $ \alpha_{i+1}\times (i)-\alpha_{i}\times (i+1), 1\leq i \leq n  $ gives us the following $ n-1 $ equations.
\begin{align}\label{a1}
\alpha_{i+1}c_{i}\eta_{i}-\alpha_{i}c_{i+1}\eta_{i+1}=0, \qquad 1\leq i \leq n-1
\end{align}
where
\begin{align}\label{eta}
\eta_{i}=\frac{a_{i}-\alpha_{i}^{2}}{\sqrt{1-a_{i}}}, \quad 1\leq i \leq n
\end{align}
Equation \eqref{a1} implies that for some ratio $ \mu$ we can write the following, 
\begin{align}\label{cb}
\begin{bmatrix}
c_{1}\eta_{1}\\
\vdots\\
c_{n}\eta_{n}
\end{bmatrix}=\mu\begin{bmatrix}
\alpha_{1}\\
\vdots\\
\alpha_{n}
\end{bmatrix}
\end{align}
Now plugging the expressions from \eqref{eta} and \eqref{cb} in any of the $ n $ equations given by \eqref{es} we get,
\begin{align}\label{main}
 \sum_{i=1}^{n}\frac{1}{1-\frac{a_{i}}{\alpha_{i}^{2}}}=1
\end{align}
It will suffice for us to prove the existance of $0<a_{i}<1, \quad 1\leq i \leq n$ such that \eqref{main} holds. From the definition of $ \eta_{i} $ given in \eqref{eta} we see that, to find each $ a_{i}, \quad 1\leq i \leq n$ we need to solve the following second order polynomial.
\begin{align}\label{poly}
a_{i}^{2}+a_{i}\alpha_{i}^{2}(\mu^{2}-2)+\alpha_{i}^{2}(\alpha_{i}^{2}-\mu^{2})=0, \quad 1\leq i \leq n 
\end{align}
If we solve equation \eqref{poly} for each $ a_{i} $ we will get a left root and a right root. Our initial conjecture is that the left root for $ a_{1} $ and right roots for $ a_{2},\dots, a_{n} $ that we get solving \eqref{poly} will give us $0<a_{i}<1, \quad 1\leq i \leq n$ that satisfy  \eqref{main}. If we can prove that our conjecture is true, then that should be the only possible solution to \eqref{main} because of the uniqueness of  solution to such convex optimization problems proved in \cite{della1982minimum}.
Plugging in the left root for $ a_{1} $, right roots for $ a_{2},\dots, a_{n} $ in \eqref{main} gives us the following equation.
\begin{align}\label{16}
1+\frac{1}{2}\sum_{i=1}^{n}\frac{\alpha_{i}^{2}}{1-\alpha_{i}^{2}}=&\frac{|\alpha_{1}|}{\sqrt{1-\alpha_{1}^{2}}}\sqrt{\frac{1}{4}\frac{\alpha_{1}^{2}}{1-\alpha_{1}^{2}}+\frac{1}{\mu^{2}}}\notag\\&-\sum_{i=2}^{n}\frac{|\alpha_{i}|}{\sqrt{1-\alpha_{i}^{2}}}\sqrt{\frac{1}{4}\frac{\alpha_{i}^{i}}{1-\alpha_{i}^{2}}+\frac{1}{\mu^{2}}}
\end{align}
We define
\begin{align}\label{xi}
X_{i}=\sqrt{\frac{1}{4}+\frac{\frac{1}{\mu^{2}}}{\frac{\alpha_{i}^{2}}{1-\alpha_{i}^{2}}}}=\sqrt{\frac{1}{4}+\frac{\frac{1}{\mu^{2}}}{\theta_{i}^{2}}}, \quad i=1,2,\dots, n
\end{align}
Under these newly defined $ X_{i} $s \eqref{16} becomes,
\begin{align}\label{plane}
 &\theta_{1}^{2}X_{1}-\sum_{i=2}^{n}\theta_{i}^{2}X_{i}=1+\frac{1}{2}\sum_{i=1}^{n}\theta_{i}^{2}
 \end{align}
 And using the definition of $ X_{i}, 1\leq i \leq n $ given in \eqref{xi}, we get the following cylinders of hyperbolas.
 \begin{align}\label{hyperbolas}
 &\theta_{1}^{2}X_{1}^{2}-\theta_{i}^{2}X_{i}^{2}=\frac{1}{4}(\theta_{1}^{2}-\theta_{i}^{2}), \quad 2\leq i \leq n
 \end{align}
 
 Equations given by \eqref{hyperbolas} imply that for each value of $ X_{1} $ we get a point $ [X_{1}, X_{2}(X_{1}), \dots ,X_{n}(X_{1}) ],$ in the $ n $ dimensional space where each $ X_{i}(X_{1}), 1\leq i \leq n $ is a function of $ X_{1} $. For the range of values  of $ (-\infty <X_{1}<\infty) $ all such points together produce an $ n $ dimensional space curve. If we project this space curve on any of the two dimensional $ X_{1}-X_{i}, 2\leq i \leq n $ planes we get a hyperbola. 

 Another important to note is that, each equation given by \eqref{hyperbolas} is a cylinder of hyperbolas originated from $ X_{1}-X_{i} $ plane and projected onto $ n $ dimensional space. Each point in the space curve represents an intersection points of all $ n-1 $ cylinders of hyperbolas originated from $ X_{1}-X_{i}, 2\leq i \leq n $ planes. The next Theorem has our revised goal at this point summed up. 
  \begin{thm}\label{thm2}
 There exists an intersection point among the plane given by \eqref{plane} and the $ n-1 $ cylinders of hypberbolas given by \eqref{hyperbolas}, that satisfies $ X_{i}>\frac{1}{2}, 1\leq i \leq n $.
 \end{thm}
 
 Proving the above Theorem would mean that, there exists $0<a_{i}<1, \quad 1\leq i \leq n$ such that \eqref{main} holds, which in turn would mean the existance of an $ n-1 $ rank CMDFA solution under the dominance of vector $ \vec{\theta}$. And as we mentioned already, the uniqueness of such solution is guaranteed.  
 
 \begin{proof}[\textbf{Outline of the Proof of Theorem \ref{thm2}:}]
 Let us define the function $ G(.) $ of $ X_{1} $ as the inner product between the vectors $ [X_{1}, \dots ,X_{n}] $ and $ [\theta_{1}^{2}, \dots ,\theta_{n}^{2}]' $ where each $ X_{i}, 1\leq i\leq n$ is a  function of $ X_{1} $. Which means,
$G(X_{1})=\theta_{1}^{2}X_{1}-\sum_{i=2}^{n}\theta_{i}^{2}X_{i}(X_{1})$. So, our revised goal becomes to find the existence of such $ X_{1}>\frac{1}{2} $ for which the function of $ G(X_{1}) $ becomes $ G(X_{1})=1+\frac{1}{2}\sum_{i=1}^{n}\theta_{i}^{2} $. Detailed  functional analysis given in \cite{} help us demostrate the convex function $ G(X_{1}) $ as in Figure \ref{figgx}.
\begin{figure}
\centering
\includegraphics[scale=0.5]{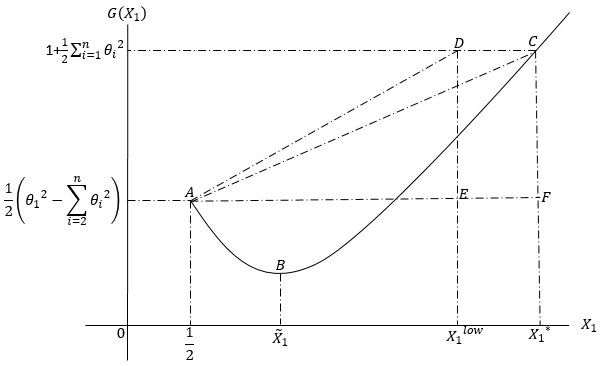}
\caption{Trend of the function $ G(X_{1} )$ against $ X_{1} $}
\label{figgx}
\end{figure}
It is straightforward to notice in Figure \ref{figgx} that $ G(X_{1}) $ is an increasing function for the values  $ X_{1}> \hat{X}_{1} $ and $G(\hat{X}_{1})<G\left(\frac{1}{2} \right)<1+\frac{1}{2}\sum_{i=1}^{n}\theta_{i}^{2}  $. Hence, there must exist $ X_{1}^{*}>\hat{X}_{1}>\frac{1}{2}  $ such that $ G(X_{1}^{*})=1+\frac{1}{2}\sum_{i=1}^{n}\theta_{i}^{2} $.
 \end{proof}
  \subsubsection{Bounds of the Solution}
 Equations \eqref{xup} and \eqref{xlow} give the expressions for the upperbound and an lowerbound to $ X_{1}^{*} $ respectively. The detailed proofs are given in \cite{}. 
 \begin{align}\label{xup}
  X_{1}^{up}=\frac{1+\frac{1}{•2}\sum_{i=1}^{n}\theta_{i}^{2}}{•\theta_{1}(\theta_{1}-\sum_{j=2}^{n}\theta_{j})}
  \end{align}
  \begin{align}
 X_{1}^{low}=\frac{1}{2}+\frac{1+\frac{1}{2}\sum_{i=2}^{n}\theta_{i}^{2}}{\theta_{1}\left(\theta_{1} -\sum_{i=2}^{n}\theta_{i}\right)}\label{xlow}
 \end{align}
 \section{Numerical Data}
 We motivated CMDFA  in terms  common information which is a function of the minimum mutual information between the observables and the latent factors.  Let $ I^{star}$, $ I^{CMDFA}$,  $ I^{up}$ and $I^{low} $   be the corresponding minimum mutual information between the latent variables and the observables for a star solution, CMDFA solution $ X_{1}^{*} $, the upperbound  $ X_{1}^{up} $  and the lowerbound $ X_{1}^{low} $. 
  \begin{figure}
\centering
\includegraphics[scale=0.5]{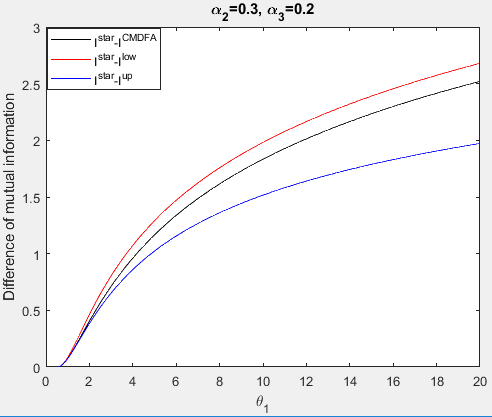}
\caption{Difference of mutual information against $ \theta_{1} $}
\label{fignumerical}
\end{figure}
  In general people tend to assume a star topology to find common information, hence any value of mutual information less than $ I^{star}$ works to our advantage. Our numerical results for a $ 3 $ dimensional case  shows that under the dominant case the star solution is not optimal.  $ I^{star}-I^{up} $ is an increasing function of $ \theta_{1} $ indicates that the lower bound of the advantage of CMDFA solution over star increases as vector $ \vec{\theta}$ becomes more and more dominant. We numerically calculated $I^{CMDFA} $ and the curve in Figure \ref{fignumerical} gives the actual advantage that CMDFA soution has over star under the dominance of $ \vec{\theta} $ whereas $ I^{star}-I^{low} $ gives an upperbound to the actual advantage of CMDFA over a star topology. The gap between $ I^{star}-I^{low} $ and $ I^{star}-I^{up} $ is gradually increasing indicating $ I^{up}-I^{low} $ is increasing with $ \theta_{1} $. The analytical justification for this numerical data is given in  \cite{}.
 \bibliography{ref}
\bibliographystyle{IEEEtran}

\end{document}